\documentclass[conference]{IEEEtran} 
\IEEEoverridecommandlockouts
\usepackage{cite}
\usepackage{amsmath,amssymb,amsfonts}
\usepackage{algorithmic}
\usepackage{graphicx}
\usepackage{textcomp}
\usepackage{xcolor}
\def\BibTeX{{\rm B\kern-.05em{\sc i\kern-.025em b}\kern-.08em
    T\kern-.1667em\lower.7ex\hbox{E}\kern-.125emX}}
\begin{document}

\title{
	Covert Communications through Imperfect Cancellation
}

\author{\IEEEauthorblockN{Daniel Chew\IEEEauthorrefmark{1}, 
		Christine Nguyen\IEEEauthorrefmark{1}, 
		Samuel Berhanu\IEEEauthorrefmark{1}, 
		Chris Baumgart\IEEEauthorrefmark{1} 
		and A. Brinton Cooper\IEEEauthorrefmark{2}}
	\IEEEauthorblockA{\IEEEauthorrefmark{1}Applied Physics Laboratory, Johns Hopkins University\\}
	\IEEEauthorblockA{\IEEEauthorrefmark{2}Electrical and Computer Engineering, Johns Hopkins University}
	}

\maketitle

\begin{abstract}
We propose a method for covert communications using an IEEE 802.11 OFDM/QAM packet as a carrier.  We show how to hide the covert message so that the transmitted signal does not violate the spectral mask specified by the standard, and we determine its impact on the OFDM packet error rate. We show conditions under which the hidden signal is not usable and those under which it can be retrieved with a usable bit error rate (BER). The hidden signal is extracted by cancellation of the OFDM signal in the covert receiver. We explore the effects of the hidden signal on OFDM parameter estimation and the covert signal BER. We conclude with an experiment using Over-The-Air recordings of 802.11 packets, inject the hidden signal, and demonstrate the effectiveness of the technique.

\end{abstract}


\section{Introduction}
Low Probability of Detection (LPD) signals are useful to reduce the probability of detection of a signal by an unauthorized observer. Typically, one of three means effect LPD communication.  The first is hiding a covert signal in additive white Gaussian noise (AWGN) such that the received signal power at an intercepting receiver is smaller than that receiver's minimum detectable signal (MDS) level \cite{Turner1991}. If the intercept receiver is merely an energy detector with an excessively wide bandwidth, detection will fail. Intercept is more likely to be successful if the receiver attempts to extract cyclostationary features from the hidden signal \cite{Spooner1994}. Second, the covert signal can be hidden in the a chosen layer of the 802.11 protocol stack \cite{Lubacz2014}, for example, by modifying the timing of the MAC layer in order to convey a pulse-width modulated message \cite{Kiyavash2013}. The trouble with these methods is that they are of low data rate. Hiding a covert signal in the physical layer offers a much higher data rate \cite{Dutta2013}, thus motivating our work. Some methods for hiding signals within OFDM symbols are explored in [6], where it is shown that a covert signal can be communicated by into an unused OFDM subcarrier, but it is seen that the information rate and the power of the covert signal are relatively small. Another physical layer scheme introduces the “dirty constellation,” by which the constellation of a signal is manipulated to a small extent so as to be indistinguishable the effects produced from ordinary wireless impairments \cite{Dutta2013} \cite{DOro2019}. For example, the constellation of a signal was perturbed just enough to convey a convert message, requiring an SNR of 24 dB in order to retrieve the covert signal \cite{Dutta2013}. Studying the injection of covert signals into OFDM in Wi-Fi, \cite{Classen2015} showed that subcarrier injection can be detected by counting the number of active subcarriers. Therefore, it is important that a covert signal not violate the spectral mask of the OFDM signal. Here, we present a novel means of transmitting and receiving a covert signal hidden inside an 802.11 OFDM packet and having a small impact on the OFDM packet while maintaining a high covert data rate. The covert signal uses a symbol rate of 156.25 kbaud and is spread using a length 64 code at a spreading rate of 10 Mcps. These rates are chosen because they are harmonics of the 20 MSPS rate of the 802.11 signal. The data rate is orders of magnitude greater than that in [6]. The covert signal is centered in the unused (null) DC bin of the OFDM packet. In our experiment, we measure the PER of the OFDM signal and then measure the BER of the covert signal, both as a function of Signal-to-Interference power Ratio (SIR). We found that the de-spreading alone was insufficient to recover the covert signal at an SIR high enough to preserve the PER of the OFDM signal. To address this problem we developed a cancellation stage in which we cancel the OFDM signal allowing the de-spreading receiver to operate in the residue. The cancellation stage provides imperfect cancellation; however, it provides sufficient suppression of the OFDM signal to recover the covert signal.  The structure of this paper is as follows: Section \ref{sec:cov_sig} describes the covert spread signal. Section \ref{sec:cov_sig_recover} considers the recovery of the covert signal using despreading alone. Section \ref{sec:OFDM_Cancellation} establishes the OFDM signal, presents our method for partially canceling the OFDM signal and demonstrates that with this cancellation we can retrieve the covert signal with significant improvements to the BER. Section \ref{sec:OTA_Test} applies the technique to Over-the-Air data and demonstrates that the cancellation technique is effective in a real world environment.

\section{Covert Signal Design}
\label{sec:cov_sig}
\begin{figure}[htbp]
	\begin{center}
		\includegraphics[width=\columnwidth]{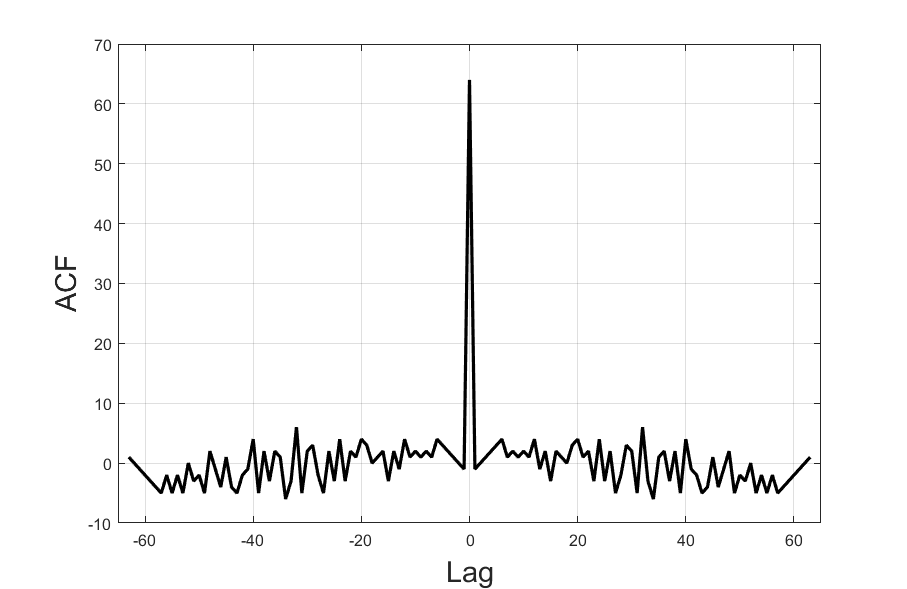}
	\end{center}
	\caption{Spreading Code ACF}
	\label{fig:SpreadingCodeACF}
\end{figure}

In this section, we develop the covert signal to be injected into an 802.11 OFDM packet. There are several unused (null) bins in the OFDM symbol as defined by the 802.11 standard \cite{80211Standard2020}. It is important that the covert signal not violate the spectral mask of the OFDM signal, as doing so would compromise the covertness. Therefore, the covert signal will be injected into the unused DC bin of the OFDM signal. If the covert signal is in the DC bin then it must not exceed the power constraints on that bin imposed by the 802.11 standard as to do so may alert a receiver the presence of the interference. 
The covert signal will be spread, and that serves multiple purposes. First, spreading the covert signal allows that signal power to be distributed across the OFDM symbol. This decreases the total detectable energy in the DC Bin.  Second, the spreading will mitigate interference and multipath. And lastly, the primary interference for the covert signal is the OFDM signal itself. The de-spreading the covert signal will spread the OFDM signal and thus reduce the total interference to the covert signal from the OFDM signal. 
For this covert signal we choose rates that are harmonics of the 20 MSPS OFDM signal for the purpose of frustrating cyclostationary analysis. The covert signal is BPSK with a symbol rate 156.25 kbaud and is spread using a length 64 code at a spreading rate of 10 Mcps. The PN sequence is derived from the polynomial $z^6+z+1$ with an initial state of [0 0 0 0 0 1] having a peak side-lobe ratio of 20.56 dB. The autocorrelation function of the spreading code is illustrated in Fig. \ref{fig:SpreadingCodeACF}.

\begin{figure}[htbp]
	\begin{center}
		\includegraphics[width=\columnwidth]{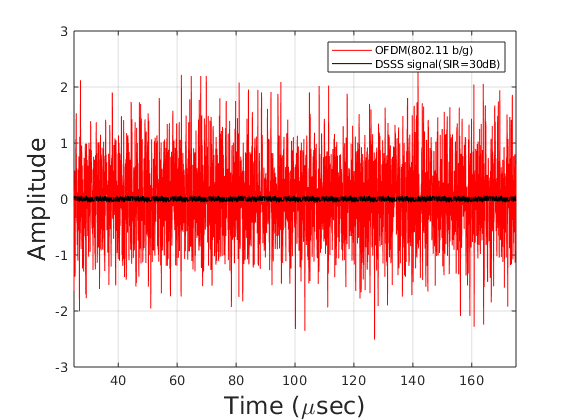}
	\end{center}
	\caption{The Covert and OFDM Signal in Time}
	\label{fig:CovertAndIncumbTime}
\end{figure}

In order to remain within the 802.11 packet, the total time that the covert signal can be transmitting is less than the total time of 802.11 packet. The relationship in time is illustrated in Fig. \ref{fig:CovertAndIncumbTime}. The figure shows the covert signal overlaid on the amplitude of the real-values of the OFDM signal. The length of the covert signal is a function of the length of the OFDM signal. Each covert symbol requires 64 chips. At 10 Mcps, this requires 6.4 microseconds per covert symbol. The OFDM signal requires 4 microseconds per OFDM symbol. Each instance of the covert signal can only have as many spread symbols as will fit inside the OFDM packet. 

\begin{figure}[htbp]
	\begin{center}
		\includegraphics[width=\columnwidth]{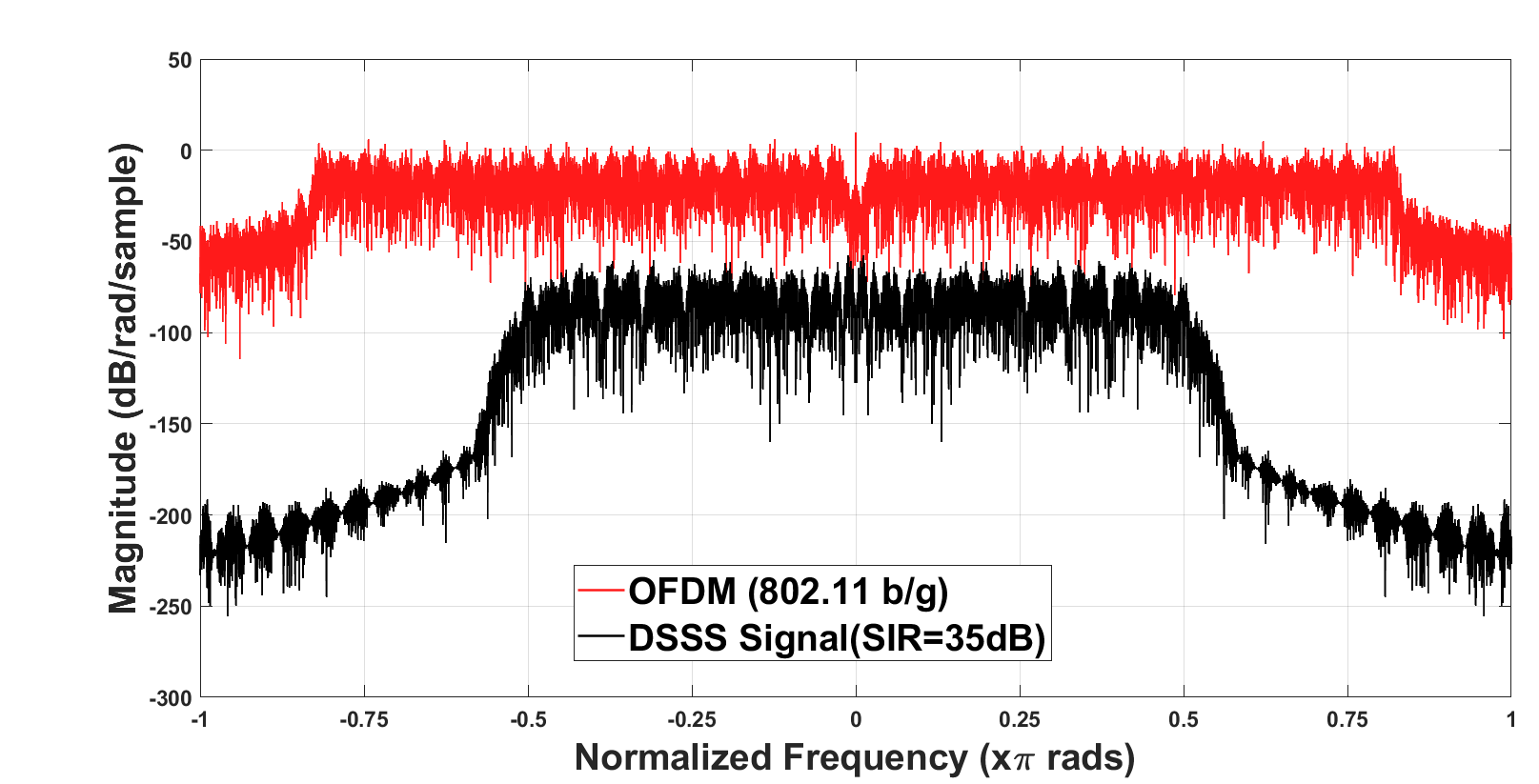}
	\end{center}
	\caption{The Covert and OFDM Signal in Frequency}
	\label{fig:CovertAndIncumbFreq}
\end{figure}

The spectrum of the OFDM signal and the covert signal are shown in Fig. \ref{fig:CovertAndIncumbFreq}. In this case the 802.11 signal is 23 dB above the noise power. The power of the covert DSSS signal is 35 dB below the OFDM signal. The covert signal adds no significant power to the sidelobes of the OFDM signal or the center bin. The effect that this covert signal has on the OFDM signal will be measured in the following sections. 

\section{System Performance Without Cancellation}
\label{sec:cov_sig_recover}
\subsection{Performance Baseline}
\label{sec:recvr_baseline}
The first step in our experiment is to establish a baseline of the performance of the 802.11 receiver in terms of PER as a function of SNR. To accomplish this, we generate synthetic packets. Each packet has a PSDU length of 1000 octets of random data. We iterate through SNR values from 0 to 25dB in steps of 0.5 dB, passing the synthetic packet through an AWGN channel. This process is repeated for MCS values 0 through 7. For each particular SNR value and MCS value, we attempt to demodulate 5000 random packets with noise, recording the packet error we receive. For this experiment, we do not add synchronization or multipath impairments. The receiver met with SNR requirements specified in the 802.11 standard for packet error rates of 10\% across MCS values.

\subsection{OFDM PER as a Function of SIR}
\begin{figure}[htbp]
	\begin{center}
		\includegraphics[width=\columnwidth]{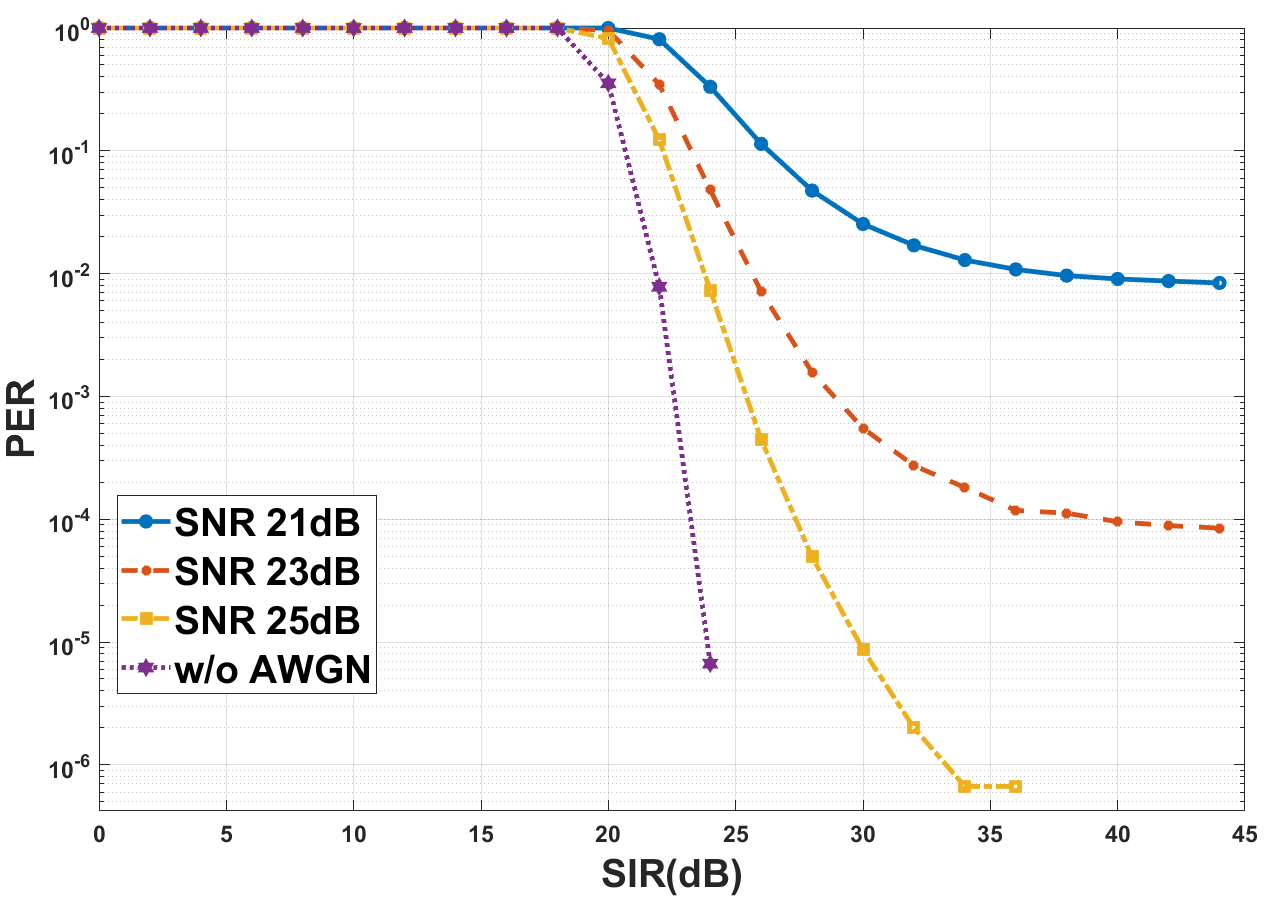}
	\end{center}
	\caption{802.11 PER at MCS 7 with a Covert Signal Present}
	\label{fig:MCS7WCovert}
\end{figure}

In an environment where an interferer tries to establish a covert channel, the predominant factor of outage for the OFDM signal can be attributed to that interferer \cite{Cardieri2010} \cite{Andrews2005}. At the receiver of the OFDM signal, packet error rates that deviate from the performance established in section \ref{sec:recvr_baseline} are good indicators for the presence of interferers. Such outage, in the case of our covert channel, can be mitigated by reducing the power spectral density (PSD) of the covert signal further after the DSSS operation. 
The reduction has the unwanted effect of lowering the covert channel’s capacity as it reduces the energy available per symbol for the covert signal. Without a mechanism to sift out the OFDM signal away from the covert channel, the processing gain achieved from the de-spreading operation does not provide a margin big enough to achieve the desired bit error performance. This is especially true if the competing objective to keep a low packet error rate for the OFDM signal is desired. This leads to a search for a SIR operating point at which one finds the lowest BER for the covert signal and the performance of the OFDM signal is within a target PER.
In order to evaluate this operating point, PER of the OFDM signal is measured over a range of SIR values. Noise power is kept constant in these measurements. Multipath and synchronization impairments were not added to the OFDM or covert signals. The problem with the covert signal interfering with the OFDM signal becomes clearer when using an MCS of 7 (64 QAM). The results of this experiment are shown in Fig. \ref{fig:MCS7WCovert}. Four different SNRs were tested: no noise, 21 dB, 23 dB, and 25 dB. The SIR for each SNR case was ranged from 0 dB to 45 dB. Note that the lowest packet error rate possible for each SNR curve is reached asymptotically indicating irreducible error rates in the presence of interferers. The results show that the SIR must be kept above 30 dB in order for the covert signal to not be the dominate contributor to the PER in the OFDM signal.  

\subsection{Covert Signal BER as a Function of SIR}
\begin{figure}[htbp]
	\begin{center}
		\includegraphics[width=\columnwidth]{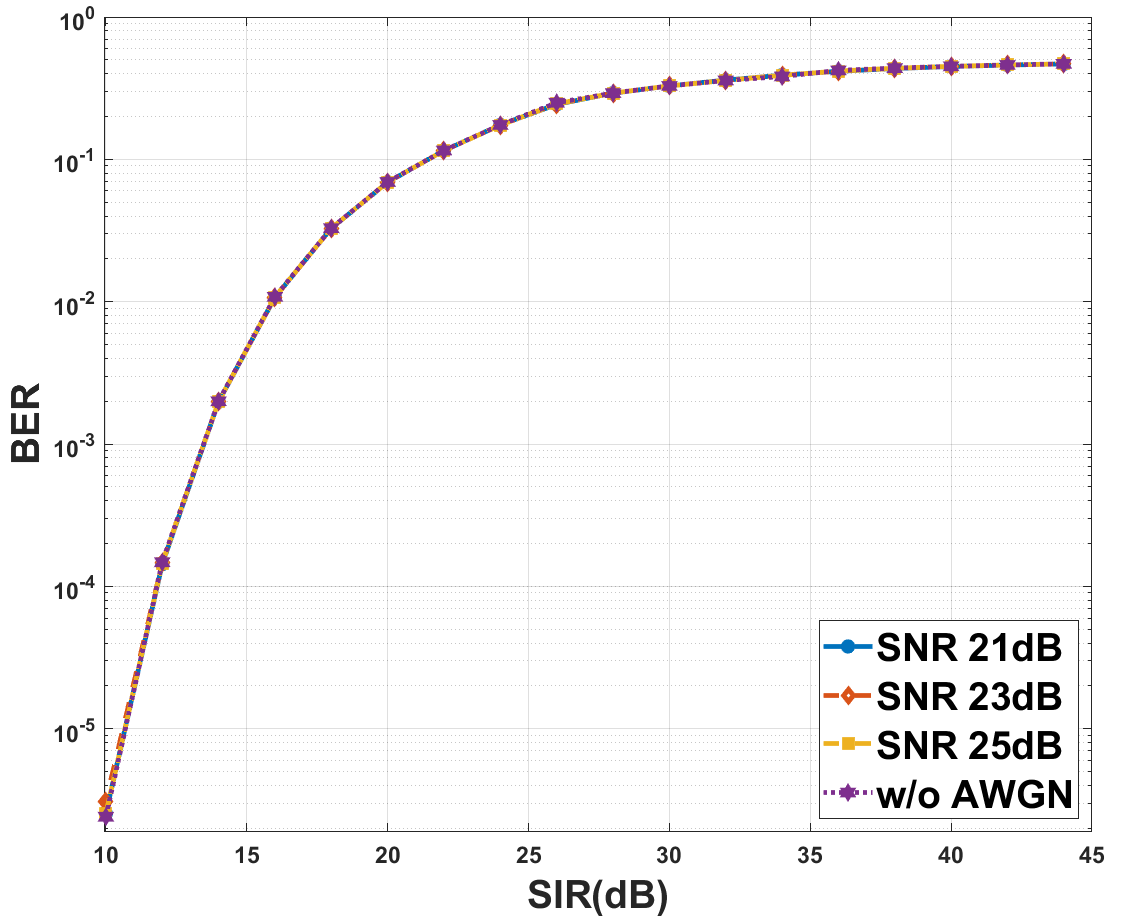}
	\end{center}
	\caption{Covert Signal BER Inside 802.11 at MCS 7}
	\label{fig:CovertBERMCS7}
\end{figure}

Fig. \ref{fig:CovertBERMCS7} depicts the bit error rate of the covert signal when the MCS is 7 (64 QAM). The four curves represent different SNR values of the OFDM packet, including one with no AWGN. The driving factor is SIR rather than SNR. The difference in BER performance under different SNR scenarios is negligible. This is a reiteration of the fact that interference from the OFDM signal and not ambient noise is the driving factor for the BER performance of the covert channel. Moreover, the SIR required to operate at $10^{-5}$ to $10^{-4}$ BER for the DSSS signal is around 12 dB. The difference between the two MCS cases is that at an MCS of 7, such SIR values significantly impact the PER performance of the 802.11 b/g receiver. One can choose a lower SIR operating point for the covert signal but the limitation as discussed earlier, is precipitated by reduced performance of the OFDM receiver.

Because the primary source of interference for the covert signal is the OFDM signal, some means of suppressing that interference must be introduced. The next section will introduce a method of partially canceling the OFDM signal.

\section{OFDM Signal Cancellation}
\label{sec:OFDM_Cancellation}

\subsection{OFDM Signal Model}
An OFDM packet is a concatenation of $N$ OFDM symbols $\vec{o}_{p}$, $p=0,1,...N-1$, where each individual symbol is created as the length-$M$ Fourier transform (IFFT) of a complex data vector $\vec{d}_{p}$ as shown in \eqref{eq:ofdmmodel} where $k$ represents discrete time for that OFDM symbol.  

\begin{equation} 
\label{eq:ofdmmodel}
o_{p} [ k ]= \sum_{m=0}^{M-1}d_{p}[m]e^{j \frac{2 \pi mk}{M}}
\end{equation}

An OFDM packet $\vec{s}$ can be represented is in \eqref{eq:packetdefinition}. The initial OFDM symbols are often defined to serve as a preamble to aid synchronization and channel estimation at the receiver.   

\begin{equation}
\label{eq:packetdefinition}
\vec{s}=\{ \vec{o}_0 \| \vec{o}_1 ... \| \vec{o}_{N-1} \}
\end{equation}

\subsection{ODFM Signal Parameter Estimation}
The demodulation process begins with receiving the preamble of the OFDM packet and performing frame synchronization. The 802.11 receiver first detects the existence of an 802.11 packet using the Long Training Field. This detection also serves as a coarse timing offset estimate. The Short Training Field is used for a fine timing offset estimate.

The OFDM packet as seen by the receiver $\vec{r}$ after timing offset correction is shown in \eqref{eq:rcvdvec} where matrix $H$ represents the multipath channel, the diagonal matrix $\Lambda_{\vec{\Theta}}$ represents the carrier frequency offset, and $\vec{n}$ noise at the receiver. There are two estimates of particular interest, those are $\widehat{\Lambda_{\Theta}}$ and $\widehat{H}$.   

\begin{equation}
\vec{r}=\Lambda_{\vec{\Theta}} H\vec{s}+\vec{n}
\label{eq:rcvdvec}
\end{equation} 

The carrier offset diagonal matrix $\Lambda_{\vec{\Theta}}$ contains phase offsets for each sample of $\vec{s}$ as shown in \eqref{eq:lambdatheta}. If the carrier frequency offset is constant then when integrated in time it will produce a phase ramp in $\Lambda_{\Theta}$. The magnitude of each element is unity. Each element on the diagonal represents a phase offset value $\Theta[k]$ shown as a phase ramp in \eqref{eq:thetak}. The carrier frequency offset can be reversed by multiplying the diagonal matrix with its conjugate. The product results in an identity matrix, $\Lambda_{\Theta}^{*} \Lambda_{\Theta}=I$.

\begin{equation}
\Lambda_{\Theta}=diag(e^{j\vec{\Theta}})
\label{eq:lambdatheta}
\end{equation}

\begin{equation}
\Theta[k]=\omega k + \phi
\label{eq:thetak}
\end{equation}

Once the first sample of the 802.11 packet has been determined, estimates are needed for $\Lambda_{\Theta}$ and $H$, those being $\widehat{\Lambda_{\Theta}}$ and $\widehat{H}$. After those parameters are estimated, the conjugate and inverse of those estimates $\widehat{\Lambda_{\Theta}}^{*}$ and $\widehat{H}^{-1}$ will be found. Because the estimates $\widehat{\Lambda_{\Theta}}$ and $H$ are not perfect representations of $\Lambda_{\Theta}$ and $H$, there will be some error when the inverse of the estimate is applied.

The first impairment to be estimated is the center frequency impairment defined in \eqref{eq:lambdatheta}. In order to correct the impairments, the estimates must be applied to the received samples. However, these impairment estimations are imperfect. The error resulting from applying the conjugate of the estimate of the carrier frequency offset is represented as the diagonal matrix $\Lambda_{\epsilon_\Theta}$ defined in \eqref{eq:CarrierEstimationError} indicating that some residual offset remains.

\begin{equation}
\widehat{\Lambda_{\vec{\Theta}}}^{*}\Lambda_{\vec{\Theta}}=\Lambda_{\epsilon_{\vec{\Theta}}}
\label{eq:CarrierEstimationError}
\end{equation}

The frequency correction estimate $\widehat{\Lambda_{\Theta}}^{*}$ is applied to the received samples $\vec{r}$ before an estimate of the channel matrix can be calculated. Therefore, the error $\Lambda_{\epsilon_{\Theta}}$ propagates into the estimation of the channel impulse response. The estimate of the channel impulse response is initially calculated using the known sequence in the 802.11 preamble. This only provides estimates for the 52 non-zero subcarriers. The channel estimate is linearly interpolated over the null-subcarriers in phase and magnitude.

If the channel impulse response has an inverse, then the inverse channel matrix $H^{-1}$ can be multiplied with $H$ and the result will be the identity matrix, $H^{-1}H=I$. The inverse of the channel estimate $\widehat{H}$ is calculated and used to equalize the received signal. The channel estimate is imperfect and therefore the equalization will be imperfect. The combined error resulting from the imperfections of both the inverse-channel estimate and the carrier frequency offset estimate is represented as $\epsilon_{H\Theta}$ as shown in \eqref{eq:TotalEstimateError}. This error trends to the identity matrix as the estimates come closer to the actual impairments.

\begin{equation}
\widehat{H}^{-1}\Lambda_{\epsilon_{\Theta}}H=\epsilon_{H\Theta}
\label{eq:TotalEstimateError}
\end{equation}

\subsection{Applying Cancellation}
\begin{figure}[htbp]
	\begin{center}
		\includegraphics[width=\columnwidth]{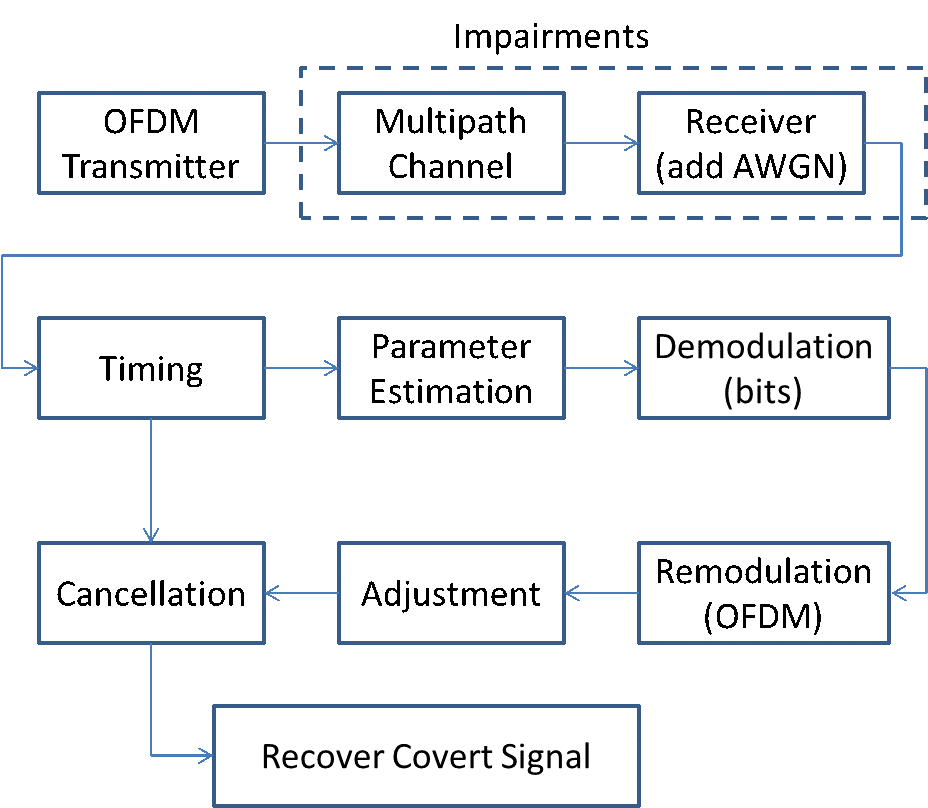}
	\end{center}
	\caption{Covert Signal Recovery with Cancellation}
	\label{fig:81png}
\end{figure}

In this section, we introduce the means by which we will add cancellation of the OFDM signal to the covert receiver. This section will build on the parameter estimation established earlier. The cancellation process is illustrated in Fig. \ref{fig:81png}. The OFDM transmitter modulates and transmits an OFDM packet. That packet passes through a multipath channel. The covert receiver estimates carrier frequency offset, channel impulse response, and demodulates the OFDM packet. The demodulated bits can be encrypted and the covert receiver has no need of decrypting these bits. The covert receiver will re-modulate the bits creating a local version of the received OFDM packet. The covert receiver will use this local copy to cancel the received OFDM signal. If the cancellation provides sufficient suppression of the OFDM signal, the covert signal can be recovered. 

Once the OFDM packet has been demodulated to bits, those bits can be re-modulated. This re-modulation produces an estimated OFDM waveform which will be represented as $\widehat{s}$. The waveform $\vec{s}$ represents the OFDM signal at the transmitter with no impairments of any kind. It is assumed that the function $g_s$ is a reproduction of the transmitter before any impairments. Therefore, any error between $\vec{s}$ and $\widehat{s}$ is attributable to bit errors, where the vector of bits used to generate $\widehat{s}$ was not the same as used to generate $\vec{s}$. The assumption that $g_s$ is a reproduction of the transmitter may not hold true if windowing was applied to the OFDM signal at the transmitter and not in $g_s$. For this paper, $g_s$ will not attempt to apply any windowing to the estimated OFDM signal $\widehat{s}$.

After the estimated corrections have been applied to the received samples, the estimated OFDM signal can be subtracted from the received samples. This subtraction results in a residue $\vec{u}$ as shown in \eqref{eq:ResidueInverseChannel}.

\begin{equation}
\vec{u}=\epsilon_{H\Theta}\vec{s}+\widehat{H}^{-1}\widehat{\Lambda_\Theta}^{-1}\vec{n}-\vec{s}
\label{eq:ResidueInverseChannel}
\end{equation}

Assuming the demodulation process saw no bit errors, and assuming function $g_s$ is a reproduction of the transmitter, then $\widehat{s}=\vec{s}$. In this case, the residue reduces to \eqref{eq:ResidueInverseChannelreduced}.

\begin{equation}
\vec{u}=(\epsilon_{H\Theta}-I)\vec{s}+\widehat{H}^{-1}\widehat{\Lambda_\Theta}^{-1}\vec{n}
\label{eq:ResidueInverseChannelreduced}
\end{equation}

The signal-error-term of the residue is $\epsilon_{(H\Theta-I)\vec{s}}$. As the estimation functions improve, $\epsilon_{H\Theta}\rightarrow I$. As that happens, the signal-error term goes to zero. The noise-term of the residue is $\widehat{H}^{-1}\widehat{\Lambda_\Theta}^{-1}\vec{n}$. The diagonal matrix $\widehat{\Lambda_\Theta}^{-1}$ will not change the autocorrelation of the noise. The inverse channel response $\widehat{H}^{-1}$ may shape the noise in spectrum meaning the autocorrelation is no longer an impulse and therefore samples of noise are no longer independent.

Shaping the noise may create problems when demodulating the covert signal. Therefore, an alternate approach to cancellation is used. Instead of applying the inverse estimated channel to the received samples, the estimate channel is applied to the signal estimate as shown in \eqref{eq:ResidueNoInverse}. This way, the noise is unfiltered.

\begin{equation}
\vec{u}=\Lambda_\Theta H\vec{s}+\vec{n}-\widehat{\Lambda_\Theta}\widehat{H}\widehat{s}
\label{eq:ResidueNoInverse}
\end{equation}

Using the same assumption about the transmitter and bit errors, the residue reduces to \eqref{eq:ResidueNoInversereduced}.

\begin{equation}
\vec{u}=(\Lambda_\Theta H-\widehat{\Lambda_\Theta}\widehat{H})\vec{s}+\vec{n}
\label{eq:ResidueNoInversereduced}
\end{equation}

The signal-error-term of the residue is $(\Lambda_\Theta H-\widehat{\Lambda_\Theta}\widehat{H})\vec{s}$. As the estimation functions improve, the estimated parameters approach the actual parameters, $\widehat{\Lambda_\Theta}\widehat{H}\rightarrow \Lambda_\Theta H$. As that happens, the signal-error term goes to zero. This would leave only the noise term in the residue. We do not expect that our signal parameters will be perfect. The cancellation technique is expected to be imperfect and we expect some remainder of the OFDM signal to be present in the residue. The goal is to estimate a sufficient number of signal parameters to create a copy of the OFDM with enough accuracy to provide significant suppression of the OFDM signal.

\section{System Improvement With Cancellation}
\label{sec:cov_sig_recover2}
For this experiment, we recover the spread signal after cancellation as described in section \ref{sec:OFDM_Cancellation}. Estimates for carrier frequency offset and the channel impulse response are created as part of the demodulation process. However, we found that the estimates of the channel impulse response and carrier frequency offset can be improved by comparing the remodulated signal to the original received signal. We adjust the power of the remodulated signal $\widehat{s}$ to match the received signal $\vec{r}$. We also compare the phase difference between $\widehat{s}$ and $\vec{r}$ and rotate $\widehat{s}$ with a phase (not frequency) offset to lower the mean difference in phase between the two signals.

The MCS of the packet is 7 (64 QAM). After cancellation, the ratio of the power of the residue and the OFDM signal represents the suppression of the OFDM signal, and that is calculated to be -20 dB. Given that the noise is 23 dB below the OFDM signal, and the residue is 20 dB below the OFDM signal, we conclude that there remains some OFDM signal power in the residue. That some OFDM signal power remains is to be expected, as the cancellation is imperfect. If the cancellation were perfect, then the residue would contain only noise. 

\begin{figure}[htbp]
	\begin{center}
		\includegraphics[width=\columnwidth]{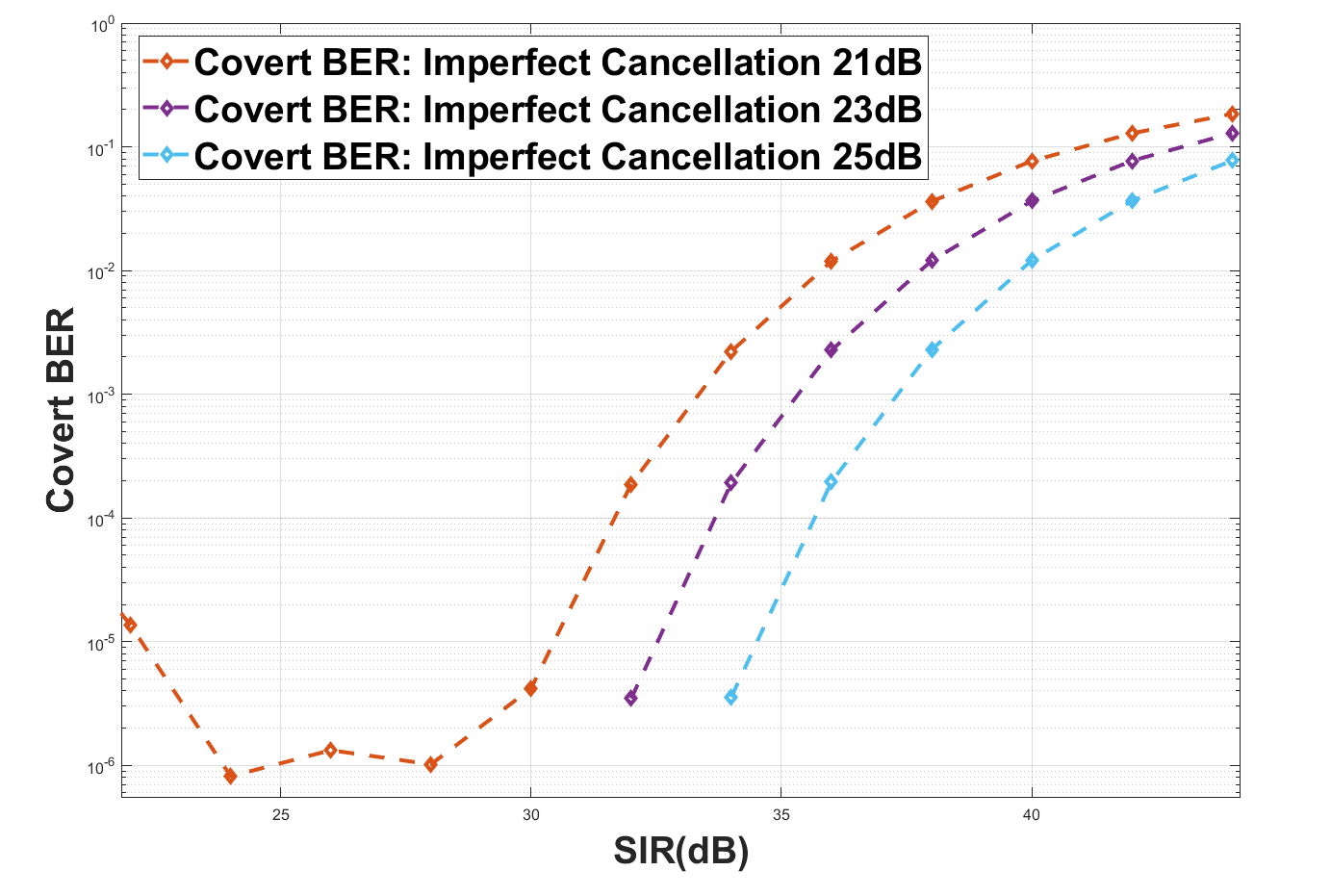}
	\end{center}
	\caption{Covert BER Improvement}
	\label{fig:CovertBERImprove}
\end{figure}

The second part of this experiment injects the covert signal into the OFDM signal before any parameter estimation is performed. The power of the covert signal is varied relative to the power of the OFDM signal, and this varies SIR and the results are plotted in Fig. \ref{fig:CovertBERImprove}. The covert signal BER dramatically improves when aided by the cancellation algorithm. The range of SIR values in which the covert signal can operate exceeds the requirements established in the analysis. Note that as SIR decreases below 24 dB, the BER of the covert signal begins to increase. This is due to the covert signal interfering with the OFDM signal so severely that it causes bit errors and thus frustrates the demodulation process illustrated in Fig. \ref {fig:81png}. 

\section{OTA Experiment}
\label{sec:OTA_Test}
For this experiment, we made complex baseband recordings of OTA Wi-Fi packets with an MCS of 7 at 40 MSPS. The Wi-Fi signals in these recordings exhibit real-world multipath channels and real-world carrier frequency offsets. Estimates for carrier frequency offset and the channel impulse response are created as part of the demodulation process. However, we found that the estimates of the channel impulse response and carrier frequency offset can be improved by comparing the remodulated signal to the original received signal. We adjust the power of the remodulated signal $\widehat{s}$ to match the received signal $\vec{r}$. We also compare the phase difference between $\widehat{s}$ and $\vec{r}$ and rotate $\widehat{s}$ with a phase (not frequency) offset to lower the mean difference in phase between the two signals.

\begin{figure}[htbp]
	\begin{center}
		\includegraphics[width=\columnwidth]{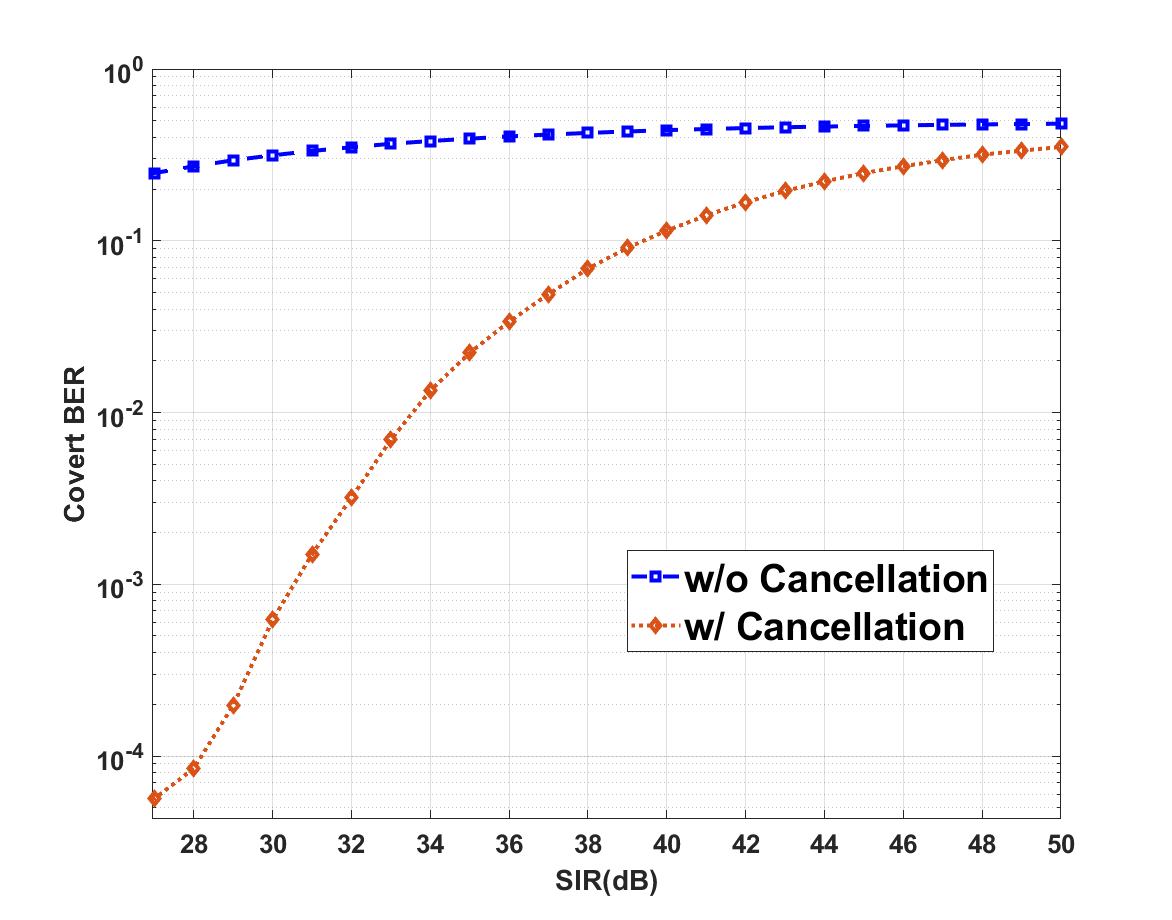}
	\end{center}
	\caption{Covert BER using OTA Data}
	\label{fig:92png}
\end{figure}

We injected the covert signal into the OTA Wi-Fi recordings. The SNR estimates of the OTA packets varied from 29 to 31 dB.  The covert signal is not perfectly aligned with the Wi-Fi packet as the Wi-Fi packet has a carrier frequency offset. Additionally, the covert signal is not impaired with a multipath channel whereas the OTA Wi-Fi Packet does exhibit that impairment. This is to say that the covert signal shares neither a frequency reference nor a multipath channel with the incumbent signal. The covert signal is injected before any parameter estimation is performed. The power of the covert signal is varied relative to the power of the OTA Wi-Fi signal, and this varies SIR. We measure the BER performance of the covert signal with and without cancellation as a function of SIR. We ran 35427 covert bits underneath 961 recorded OTA packet over a range of SIR values. The results are plotted in Fig. \ref{fig:92png}. The average cancellation was 18.5 dB, which is lower than observed in simulation. The covert signal BER dramatically improves when aided by the cancellation algorithm. The range of SIR values in which the covert signal can operate exceeds the requirements established in analysis.

\section{Conclusion}
\label{sec:conclusion}
In this paper, we have measured the impact of a DSSS spread covert signal on an OFDM signal. We have shown that SIR, not SNR, is the primary limitation of the covert signal sharing the link with the OFDM signal. When high modulation orders are used in the OFDM signal, the covert signal must operate at high SIR values. The high SIR value means that the covert signal will be pushed deep into the noise at the 20 MHz 802.11 receiver. At that noise power, it is demonstrated that the covert link will experience difficulties. One means to mitigate those difficulties is to reduce the data rate of the covert signal. At a lower data rate, the covert signal can enjoy a smaller bandwidth and thus lower noise power. In this paper, we present an alternative to reducing the bandwidth of the covert signal. We implement imperfect cancellation of the OFDM signal. We demonstrate this imperfect cancellation is sufficient to recover a 156.25 kbaud BPSK signal spread at 10 Mcps from underneath an OFDM signal using 64 QAM both in simulation and when using OTA data. We observed that the cancellation on OTA data is lower than that of our simulation, and we attribute that to signal parameters not estimated in our signal model. Our covert signal is capable of operating at SIR values higher than 30 dB, which maintains the throughput of the OFDM signal.

\bibliographystyle{IEEEtran}
\bibliography{IEEEabrv,CovertCommunicationsImperfectCancellation}

\end{document}